\documentclass[vecphys]{svmult}
\usepackage{graphicx}        

\begin{document}
\title*{Evolution of helicity in NOAA 10923 over three consecutive solar rotations}
\author{Sanjiv Kumar Tiwari\thanks{e-mail: stiwari@prl.res.in},
Jayant Joshi, Sanjay Gosain \\ and P. Venkatakrishnan}
\authorrunning{S.Tiwari et al.}
\institute{Udaipur Solar Observatory,
         Physical Research Laboratory, \\
P. Box - 198, Dewali, Bari Road, Udaipur-313 001, \\ Rajasthan, India}

%
%
\maketitle

\begin{abstract}
  We have studied the evolution of magnetic helicity and chirality in
  an active region over three consecutive solar rotations. The region
  when it first appeared was named NOAA10923 and in subsequent
  rotations it was numbered NOAA 10930, 10935 and 10941. We compare
  the chirality of these regions at photospheric, chromospheric and
  coronal heights. The observations used for photospheric and
  chromospheric heights are taken from Solar Vector Magnetograph (SVM)
  and H-$\alpha$ imaging telescope of Udaipur Solar Observatory (USO),
  respectively. We discuss the chirality of the sunspots and
  associated H-$\alpha$ filaments in these regions. We find that the
  twistedness of superpenumbral filaments is maintained in the
  photospheric transverse field vectors also. We also compare the
  chirality at photospheric and chromospheric heights with the
  chirality of the associated coronal loops, as observed from the
  HINODE X-Ray Telescope.
\end{abstract}


\section{Introduction}
Magnetic fields exhibit chirality and is observed in most of the solar
features like filament channels, filaments, sunspots, coronal loops,
coronal X-Ray arcades and interplanetary magnetic clouds (IMCs)
(\cite{seehafer:90}, \cite{pevtsov:95}, \cite{martin:98} and the references
therein). First of all G.E. Hale in 1925 (\cite{hale:25}) observed vortices in
H-$\alpha$ around sunspots and he called these features as 'sunspot
whirls'. He investigated the data extending over three solar cycles
and found no relationship between the direction of these vortices and
the polarity of the sunspots. Also, he found no reversal of the whirl
direction together with the general reversal of the sunspot polarities
with cycle. He found that about 80 \% of sunspot whirls have
counterclockwise orientation in the northern hemisphere and clockwise
in the southern hemisphere, now known as the helicity hemispheric
rule. Richardson (\cite{richardson:41}) confirmed Hale's results after doing the same
type of investigation with the data for four solar cycles. Seehafer
(\cite{seehafer:90}) also found that the negative helicity is dominant in the
northern hemisphere and positive in the southern hemisphere.  This is
known as the hemispheric helicity rule and is independent of the solar
cycle. Since 90s the subject is highly taken into account by most of
the researchers in the field. Pevtsov, Canfield and Metcalf (\cite{pevtsov:95})
demonstrated the existence of chirality in the active regions, after
analyzing vector magnetograms from Mees Solar Observatory. Although,
the entire active region may not show the same type of chirality
everywhere yet a dominant sense of chirality can be found for most of
the active regions.

The chirality of the H-$\alpha$ filament can be directly recognized by
looking at the filament barbs. If the orientation of the barbs is
counterclockwise the chirality of the filament is known to be dextral
and if it is clockwise the chirality is sinistral. Martin (\cite{martin:06})
mentioned that the chirality of the solar features can be used for
resolving 180 degree azimuthal ambiguity in the solar vector magnetic
field.  It is believed that there is one-to-one correspondence between
the filament chirality and the magnetic helicity sign. A right-handed
twist and a clockwise rotation of the loops, when viewed from the
above implies positive helicity or chirality and vice-versa. The
magnetic helicity is a quantitative measure of the chiral properties
of the solar magnetic structures (\cite{berger:84}, \cite{berger:99}).
It is given by a volume integral over the scalar product of the
magnetic field B and its vector potential A.
\begin{equation}\
   H = \int\mathbf{A}\cdot\mathbf{B} dV .
\end{equation}
It is well known that the vector potential A is not unique. Thus the
helicity can't immediately be calculated from the above given
equation. Seehafer (\cite{seehafer:90}) pointed out that the helicity of magnetic
field can best be characterized by force-free parameter alpha also
known as the helicity parameter. The force free condition is given as
            \begin{equation}
           \nabla\times\mathbf{B}=\alpha\mathbf{B}.
            \end{equation}
Taking the z-component of the magnetic field
\begin{equation}
\alpha=(\nabla\times\mathbf{B})_{z}/\mathbf{B}_{z}
\end{equation}

The magnetic helicity density can be given as
\begin{equation}
    H_{m}=B^{2}/\alpha
\end{equation}
but except for potential fields.\\
And the current helicity density will be given in terms of alpha as
 \begin{equation}
 H_{c}=B^{2}\alpha
 \end{equation}

 In this paper, we identify the chirality of a sunspot in an active
 region named NOAA 10923 when it first appeared and NOAA 10930, 10935,
 10941 in successive rotations. The associated H-$\alpha$ filaments in
 the three consecutive solar rotations were obtained from the Udaipur
 Solar Observatory (USO) high resolution H-$\alpha$ data. We have
 calculated the helicity parameter of the sunspots NOAA 10935 and NOAA
 10941 and we found that the value of helicity parameter has increased
 after one solar rotation. These active regions are not following the
 helicity hemispheric rule. There are theories (\cite{choudhuri:04},
 \cite{zhang:06},\cite{chaterjee:06}) which discuss about the 'wrong' sign
 of helicity in the beginning of the solar cycle. Sokoloff et al
 (\cite{sokoloff:06}) observes a significant excess of active regions with the
 'wrong' sign of helicity just at the beginning of the cycle. But
 there is no discussion found about the 'wrong' sign of helicity
 during the end of the solar cycle. The sign of helicity is supposed
 to follow the helicity hemispheric rule which our result doesn't
 show. We compare the helicity of the active regions with the sign of
 associated sigmoids obtained from the Hinode.

\section{Data and Instruments used }

We use high resolution H-$\alpha$ images taken from Udaipur Solar
Observatory (USO) from Spar Telescope and the vector magnetograms from
the USO Solar Vector Magnetograph (SVM) (\cite{gosain:05}, \cite{gosain:06}). The
Spar Telescope uses 1392$\times$1024 ccd with the pixel resolution of
0.395 arcsec and the SVM has 1024$\times$1024 ccd with pixel
resolution of 0.98 arcsec.  Our H-$\alpha$ observations, after the
monsoon break, started on 23rd November 2006. So we don't have USO
H-$\alpha$ image of the NOAA 10923 in its first appearance. After one
rotation we observe the same sunspot in the name of NOAA 10930 on Dec
11 2006.  And in other two consecutive rotations we have observed the
same sunspot in the name of NOAA 10935 on 09 Jan 2007 and NOAA 10941
on 06 Feb 2007. We use the available vector magnetograms of NOAA 10935
(09 Jan 07) and NOAA 10941 (06 Feb 07) and compare the vectors with
the whirls of the sunspot images taken in H-$\alpha$ wavelength.

The Spar Telescope has f/15 doublet lens with focal length of 2.25
meters and objective 0.15 meters.  It uses a H-$\alpha$ Halle lyot
type filter with FWHM of 500 m$\AA$ operating at the wavelength of
6563 $\AA$. The telescope utilizes a 1392$\times$1024 CCD with the
pixel size of 6.45 $\mu$m. The pixel resolution of the CCD is 0.395
arc-sec and the field of view it covers is 9 arc-min $\times$ 7
arc-min. The H-alpha images of the active regions NOAA 10930, NOAA
10935 and NOAA 10941 are shown in the Figure 1.

\begin{figure}
\centering
\includegraphics[height=8.2truecm]{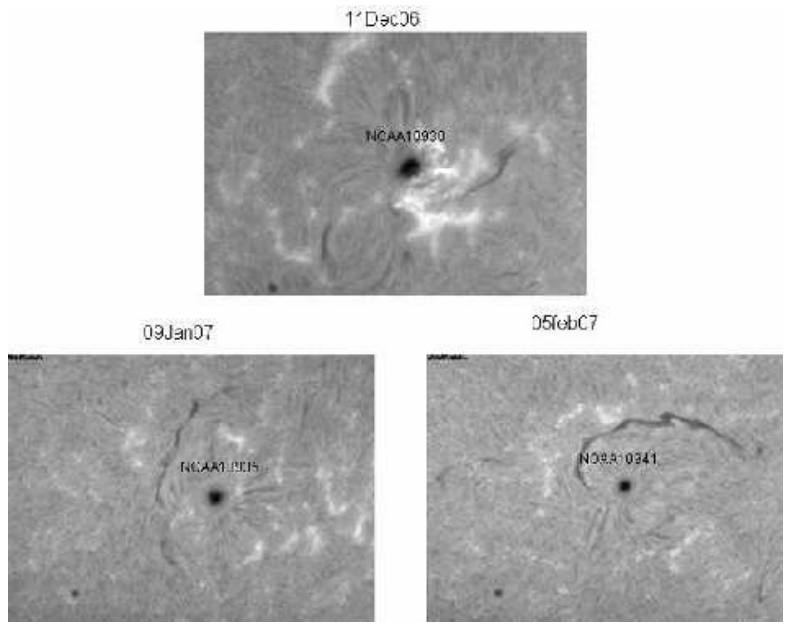}
\caption{USO H-alpha images }
\label{fig1}
\end{figure}

\begin{figure}
\centering
\includegraphics[height=5.5truecm]{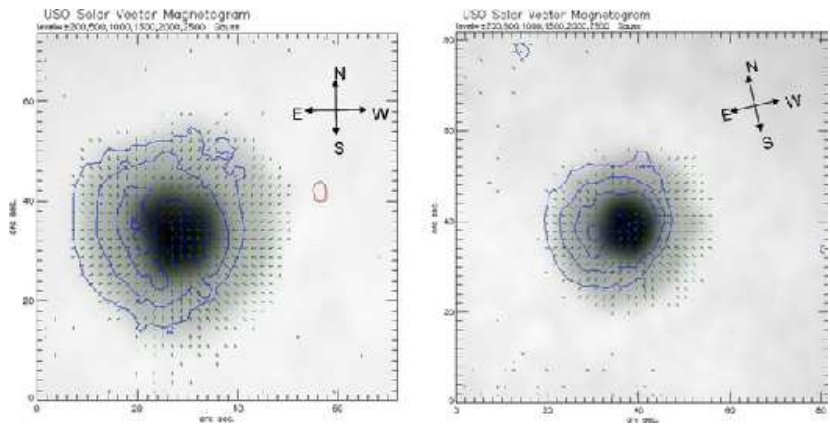}
\caption{ USO solar vector magnetograms
  (09Jan07 and 06Feb07 respectively) }
\label{fig2}
\end{figure}

\begin{figure}
\centering
\includegraphics[height=6.2truecm]{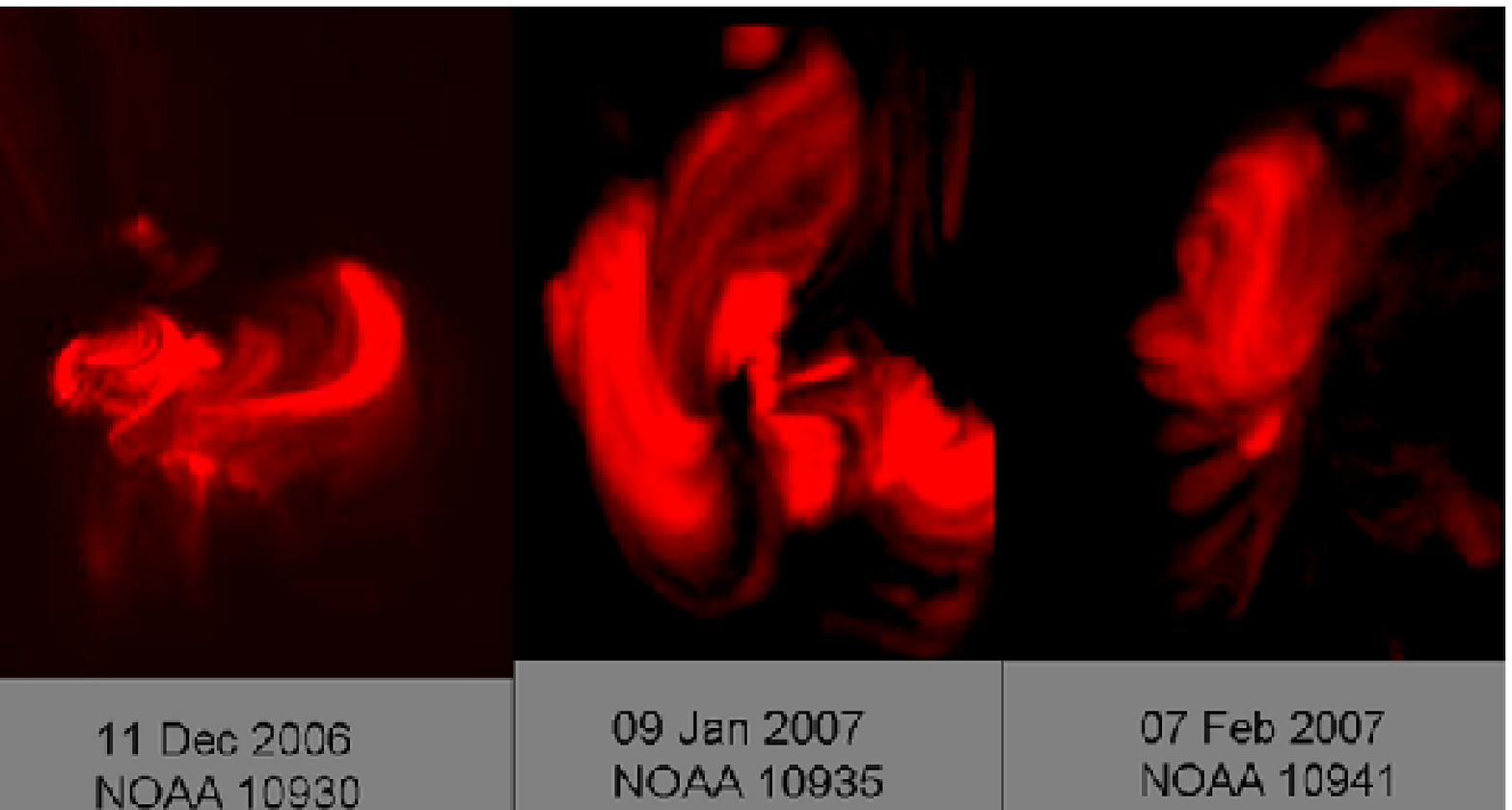}
\caption{ XRT (Hinode) data reverse-S sigmoid
  showing the dextral chirality in comparison with that of
  photospheric and chromospheric data. }
\label{fig3}
\end{figure}

The magnetograms are taken from Solar Vector Magnetograph (SVM), which
has recently become operational at USO. SVM is basically an instrument
which makes two-dimensional spatial maps of solar active regions in
the Zeeman induced polarized light of the solar spectral lines. SVM
has the following main components : a Schmidt-Cassegrain telescope
tube, rotating wave-plate polarimeter, tunable narrow-band Fabry-Perot
filter, calcite analyzer (Savart plate) and a cooled CCD camera. The
primary imaging is done by using a Celestron C-8 (TM)
Schmidt-Cassegrain telescope of 8 inch aperture. The focal length of
the telescope is 2032 mm and the resulting output beam is a f/10
beam. The telescope has a pre-filter in front with a 15nm pass-band
centered at 630nm wavelength. A circular aperture of 2 arc-min
diameter selects the field of view (FOV) at the prime focus. This FOV
is then modulated by the rotating waveplates of the polarimeter. The
modulated beam is now collimated by a 180mm focal length lens. This
modulated and collimated beam now enters the Fabry-Perot etalon and
order sorting pre-filter. Now the re-imaging lens makes the image on
the CCD camera. Just before the CCD camera a combination of two
crossed calcite beam-displacing crystals is placed for the analysis of
polarization. So we get two orthogonal polarized images of the
selected FOV onto the CCD camera. The vector magnetograms of the two
active regions NOAA 10935 and NOAA 10941 are shown in the Figure - 2.
We have taken the Hinode (XRT)(\cite{Golub:07}) data for looking
at the sigmoidal structure of corresponding active regions.

\section{Analysis }

We observed the sunspot NOAA 10923 (Nov 06) which sustains in the three
consecutive solar rotation in the name of NOAA 10930 (Dec 06), NOAA
10935 (Jan 07) and NOAA 10941 (Feb 07). There was no major activity
associated with the active region NOAA 10923.  Looking at the
NOAA 10930 in H-Alpha (Fig(1)) we ensure that there is no particular
orientation of the whirls and we can't recognize the helicity of the
sunspot. There was no filament seen associated with this active
region. There were X3.4 (at 02:00 UT) and X1.5 (at 21:07 UT) class flares as
well as strong CMEs associated with the active region NOAA 10930 on 13
December 2006.  After next rotation the same sunspot is found in the
name of NOAA 10935 (Jan 07) (Fig(1)). We now observe the whirls with
counter-clockwise orientation are dominating. A filament associated
with the active region is observed. The end of this active region
filament is curving towards the sunspot with counter - clockwise
whirls and according to Rust and Martin (\cite{rust:94}) it should (not
necessarily) be dextral.  Still in the image here the orientation of
the filament barbs are not clearly recognized to be dextral. There was
not any major event associated with this active region.
\begin{figure}
\centering
\includegraphics[height=9.1truecm]{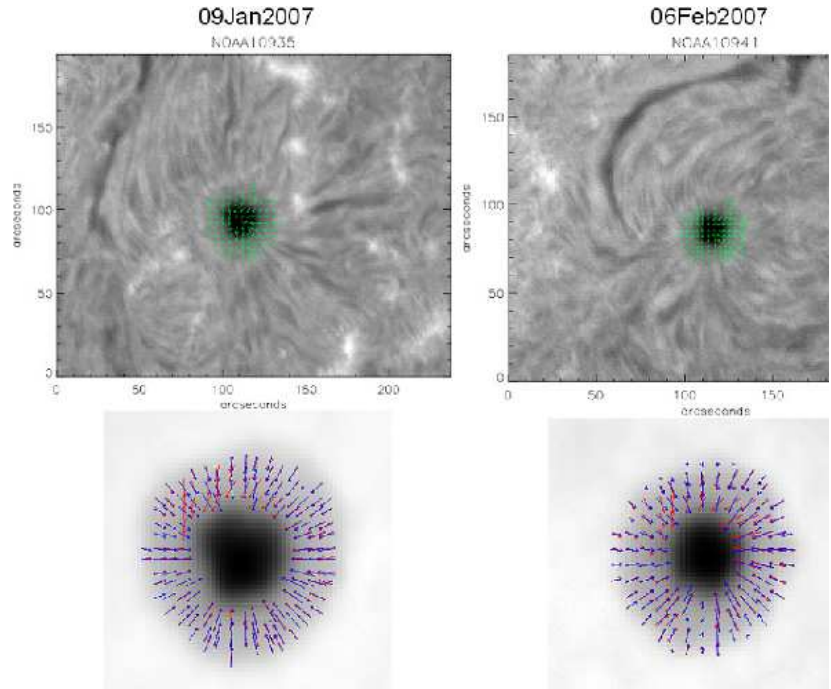}
\caption{ Plot of vector magnetic fields
    of the sunspots upon the respective H-alpha images. In the lower
    part of the image blue arrows show the radial direction and the
    red arrows show the actual vector field direction. We can see the
    shear of the field.}
\label{fig4}
\end{figure}

In the next rotation we are able to recognize clearly the orientation
of the filament barbs. The active region NOAA 10941 (Feb 07) has come
in its third consecutive rotation of the sun. It has the barbs with
counter - clockwise orientation close to filament and this type of
chirality is dominating in the whole active region. The associated
filament is dextral is clear now. No event was observed associated
with this active region.  We calculated the helicity parameter for the
sunspots NOAA 10935 and NOAA 10941 which comes out to be -1.1$\pm 0.12
\times 10^{-9}$~$m^{-1}$ and -$3.77 \pm2.14 \times 10^{-8} m^{-1}$
respectively. The value of helicity has increased after next rotation
which can also verified by directly looking at the H-alpha images. The
helicity sign doesn't support the helicity hemispheric rule. Both the
active regions are found in the southern hemisphere and should bear
positive chirality.  For the calculation of helicity we have
calculated the helicity parameter alpha best which will give one value
of the alpha for the whole sunspot instead of the value at each
pixel. It reduces the noise in the data.  We use our SVM data to plot
the transverse field upon the associated H-$\alpha$ image. First of
all, the H-$\alpha$ data is re-binned according to the size of SVM
data and then the related vector field is plotted over the H-$\alpha$
image.  The 180 degree azimuthal ambiguity has been resolved by using
acute angle method. We can see in Figure - 2 that the direction of
H-$\alpha$ super-penumbral whirls show the same chirality as the
photospheric transverse field vectors. So, by combining the
photospheric and chromospheric data one can use the method of
chirality to resolve 180 degree azimuthal ambiguity.  Also by looking
at the HINODE (XRT) data we find the reverse-S structure in the
sigmoids associated with the active regions.


\section{Conclusion and Discussion }

We find that the active region NOAA 10923 in its different appearances
doesn't follow the hemispheric helicity rule, neither its associated
filaments. But their association in the terms of chirality follow the
well known result of Rust and Martin (\cite{rust:94}). We calculated the
helicity parameter which was found to be negative as expected from the sunspot
with dextral whirls. The helicity increases in the last appearance
but there was no major activity observed associated with the active
region in its last appearance. Hinode (XRT) images also show the same
type of chirality in the associated sigmoids. Thus the sign of
helicity derived from photosphere, chromosphere and corona are
strongly correlated.  Figure - 3 shows the vector fields plotted over
the associated active regions in the chromospheric H-$\alpha$
observations. We find the good matching of the vectors with the
super-penumbral whirls of the active regions. By combining the
photospheric and chromospheric data the axial field direction at
neutral line can be derived and this method can be used to resolve 180
degree azimuthal ambiguity. Sara F. Martin et al (\cite{martin:06}) has already
mentioned that the 180 degree azimuthal ambiguity can be resolved by
using this method of chirality.\\

\paragraph{\bf Acknowledgement }
 Hinode is a Japanese mission developed and launched by ISAS/JAXA, collaborating with NAOJ
  as a domestic partner, NASA and STFC (UK) as international partners. Scientific operation
   of the Hinode mission is conducted by the Hinode science team organized at ISAS/JAXA.
    This team mainly consists of scientists from institutes in the partner countries.
    Support for the post-launch operation is provided by JAXA and NAOJ (Japan),
     STFC (U.K.), NASA (U.S.A.), ESA, and NSC (Norway). One of us (Jayant Joshi)
     acknowledge financial support under ISRO / CAWSES -  India programme.




\begin{thebibliography}{99.}

\bibitem{seehafer:90} Seehafer, N., 1990,
         \newblock{\ Solar Phys.}, {\bf 125}, 219.

\bibitem{pevtsov:95} Pevtsov, A., Canfield, R. C. and Metcalf, T. R., 1995,
        \newblock{\ ApJ}, {\bf 440}, L109

\bibitem{martin:98} Martin, S. F., 1998,
       \newblock{\  ASP Conference Series}, {\bf 150}, 419.

\bibitem{hale:25} Hale,G. E., 1925,
       \newblock{\ Publ. Astron. Soc. Pacific}, {\bf 37}, 268.

\bibitem{richardson:41} Richardson, R. S., 1941,
        \newblock{\ Astrophys. J.}, {\bf 93}, 24

\bibitem{martin:06} Martin, S. F., Lin, Y., Engvold, O., 2006,
        \newblock{\ American Astronomocal Society, SPD Meeting}, {\bf 37}, 129

\bibitem{berger:84} Berger, M.~A. and Field, G. B., 1984,
           \newblock {\ J. Fluid Mech.}, {\bf 147}, 133.

\bibitem{berger:99} Berger, M.~A., 1999,
           \newblock {\ Magnetic Helicity in Space and Laboratory Plasmas, Geophys. Monograph}, {Vol. \bf 111},
           {American Geophysical Union}, p.1.

\bibitem{choudhuri:04} Choudhuri, A. R. et al., 2004,
      \newblock {\ Astrophys. J.}, {\bf 615}, L57.

\bibitem{zhang:06} Zhang, H. et al., 2006
        \newblock{\ MNRAS }, {\bf 365}, 276 (Paper I)

\bibitem{chaterjee:06} Chaterjee, P., 2006,
        \newblock {\ J. Astron. Astrophys.}, {\bf 27}, 89.

\bibitem{sokoloff:06} Sokoloff, D. , 2006,
        \newblock{\ Astronomische Nachrichten}, {\bf 327, Issue 9}, 876

\bibitem{gosain:05} Gosain, S., Venkatakrishnan, P. and Venugopalan, K.; 2005,
       \newblock {\ Exp.Astron.},{\bf 18 }, 31.

\bibitem{gosain:06}Gosain, S., Venkatakrishnan, P. and Venugopalan, K., 2006,
        \newblock {\ J. Astrophys. Astron.}, {\bf 27}, 285.

\bibitem{Golub:07}Golub, L. et al.; 2007,
        \newblock {\ Solar Phys.}, {\bf 243}, 63.

\bibitem{rust:94} Rust, M. D. and Martin, S. F., 1994,
         \newblock{\ ASP Conference Series}, {\bf 68}, 337.



\end{thebibliography}
\end{document}